\documentclass[aps,prl,twocolumn,showpacs,groupedaddress]{revtex4}
\usepackage{amssymb}
\usepackage{graphicx}
\usepackage{bm}
\begin{document}
\title{Valley effects on the fractions in an ultrahigh mobility SiGe/Si/SiGe two-dimensional electron system}
\author{V.~T. Dolgopolov}
\affiliation{Institute of Solid State Physics, Chernogolovka, Moscow District 142432, Russia}
\author{M.~Yu.~Melnikov}
\affiliation{Institute of Solid State Physics, Chernogolovka, Moscow District 142432, Russia}
\author{A.~A. Shashkin}
\affiliation{Institute of Solid State Physics, Chernogolovka, Moscow District 142432, Russia}
\author{S.-H. Huang and C.~W. Liu}
\affiliation{Department of Electrical Engineering and Graduate Institute of Electronics Engineering, National Taiwan University, Taipei 106, Taiwan, and\\ National Nano Device Laboratories, Hsinchu 300, Taiwan}
\author{S.~V. Kravchenko}
\affiliation{Physics Department, Northeastern University, Boston, Massachusetts 02115, USA}
\begin{abstract}
We observe minima of the longitudinal resistance corresponding to the quantum Hall effect of composite fermions at quantum numbers $p=1$, 2, 3, 4, and 6 in an ultraclean strongly interacting bivalley SiGe/Si/SiGe two-dimensional electron system. The minima at $p=3$ disappear below a certain electron density, although the surrounding minima at $p=2$ and $p=4$ survive at significantly lower densities. Furthermore, the onset for the resistance minimum at a filling factor $\nu=3/5$ is found to be independent of the tilt angle of the magnetic field. These surprising results indicate the intersection or merging of the quantum levels of composite fermions with different valley indices, which reveals the valley effect on fractions.
\end{abstract}
\maketitle

The concept of composite fermions \cite{jain1989composite,halperin1993theory,jain1994composite,chakraborty2000electron,jain2007composite} can successfully describe the fractional quantum Hall effect with odd denominators by reducing it to the ordinary integer quantum Hall effect for composite particles. In the simplest case, the composite fermion consists of an electron and two magnetic flux quanta and moves in an effective magnetic field $B^*$ given by the difference between the external magnetic field $B$ and the field corresponding to the filling factor for electrons, $\nu=n_{\text s}hc/eB$, equal to $\nu=1/2$, where $n_{\text s}$ is the electron density. The filling factor for composite fermions, $p$, is connected to $\nu$ according to the expression $\nu=p/(2p\pm1)$. The fractional energy gap, which is predicted to be determined by the Coulomb interaction in the form $e^2/\varepsilon l_{\text B}$, corresponds to the cyclotron energy of composite fermions $\hbar\omega_{\text c}^*=\hbar eB^*/m_{\text{CF}}c$, where $\varepsilon$ is the dielectric constant, $l_{\text B}=(\hbar c/eB)^{1/2}$ is the magnetic length, and $m_{\text{CF}}$ is the effective composite fermion mass. The electron-electron interactions enter the theory \cite{jain1989composite,halperin1993theory,jain1994composite,chakraborty2000electron,jain2007composite} implicitly because a mean-field approximation is employed, assuming that the electron density fluctuations are small. The theory is confirmed by the experimental observation of a scale corresponding to the Fermi momentum of composite fermions in zero effective magnetic field at $\nu=1/2$.

The majority of the experiments on the fractional quantum Hall effect have been performed on single-valley GaAs-based heterostructures \cite{jain2007composite}. There are only a few experiments on other two-dimensional (2D) electron systems in which electrons occupy two valleys, \textit{e.g.}, in (001) SiGe heterostructures \cite{lai2004two,lu2009observation,melnikov2019quantum} and (001) AlAs quantum wells \cite{bishop2007valley,padmanabhan2009density,padmanabhan2010ferromagnetic,padmanabhan2010composite}. A significant advantage of the ultraclean bivalley 2D electron system in SiGe heterostructures (as well as the 2D hole system in GaAs/AlGaAs heterostructures) compared to the 2D electron system in GaAs/AlGaAs heterostructures is that at accessible low electron densities/weak magnetic fields, the limit can be reached where the electron interaction energy $e^2/\varepsilon l_{\text B}$ becomes much greater than the cyclotron energy $\hbar eB/mc$, where $m$ is the effective electron mass. In this case, the fractional gap can exceed some of the other spectral gaps, \textit{e.g.}, the spin or valley gaps, leading to a change in the gap systematics in the spectrum. In the bivalley 2D electron system in SiGe quantum wells, an unusual behavior has been observed in the longitudinal resistance at low densities: the dominance of the $\nu=2/5$ fraction over the $\nu=1/3$ fraction \cite{melnikov2019quantum}, which is in contrast to the results obtained on GaAs-based heterostructures and in disagreement with the hierarchy of fractions, as inferred from the concept of composite fermions. The origin of the discrepancy has not yet been explained. Notably, two additional minima in the magnetoresistance at $\nu=4/5$ and $\nu=4/11$ that are symmetric relative to $\nu=1/2$ have been recently observed in 2D electron systems in both GaAs/AlGaAs quantum wells \cite{pan2015fractional} and SiGe quantum wells \cite{dolgopolov2018fractional}. Both minima correspond to the fractional filling factor of composite fermions, $p=4/3$, which suggests a formation of the second generation of composite fermions on the basis of already existing ones, as described by the double attachment of two magnetic flux quanta \cite{dolgopolov2018fractional,park2001the}.

In this Letter, we observe minima of the longitudinal resistance corresponding to the quantum Hall effect of composite fermions at quantum numbers $p=1$, 2, 3, 4, and 6 in an ultraclean strongly interacting bivalley SiGe/Si/SiGe 2D electron system. The minima at $p=3$ disappear below a certain electron density, although the surrounding minima at $p=2$ and $p=4$ persist to significantly lower densities. This finding is strikingly similar to the effect of the disappearance of the cyclotron minima in the magnetoresistance at low electron densities in Si metal-oxide-semiconductor field-effect transistors (MOSFETs) while the spin minima survive down to appreciably lower densities \cite{kravchenko2000shubnikov}, which signifies that the cyclotron splitting becomes equal to the sum of the spin and valley splittings, and the corresponding valley sublevels merge \cite{shashkin2014merging}. Furthermore, the onset for the resistance minimum at filling factor $\nu=3/5$ is found to be independent of the tilt angle of the magnetic field, excluding the spin origin of the effect. The results point to the intersection or merging of the quantum levels of composite fermions with different valley indices, which reveals the valley effect on the fractions. In analogy with Si MOSFETs, it is very likely that the merging of the composite fermion levels with different valley indices occurs in our samples. In contrast to the simple crossing of quantum levels, the effect of the interaction-caused level merging is very similar to the band flattening at the Fermi level in strongly correlated Fermi systems, which has been indicated in a SiGe/Si/SiGe 2D electron system based on the experiment in Ref.~\cite{melnikov2017indication}. The dominance of the $\nu=2/5$ fraction over the $\nu=1/3$ fraction, observed earlier in the magnetoresistance at low densities in this electron system, may also be related to valley effects. The observed experimental effects are surprising and warrant further theoretical consideration.

\begin{figure}
\scalebox{0.43}{\includegraphics[angle=0]{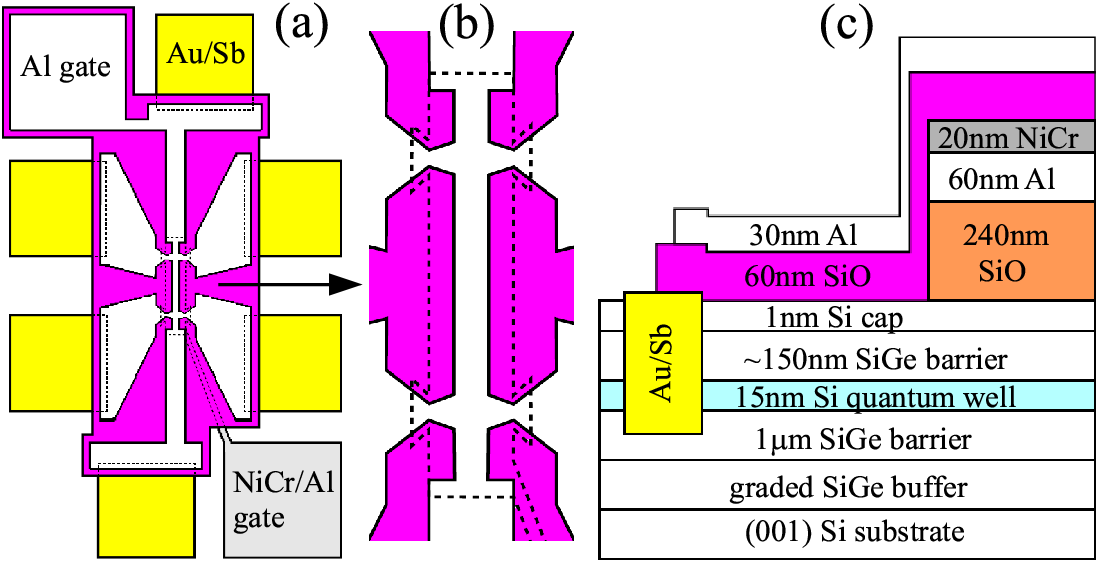}}
\caption{(a) Schematic top view on the sample with two independent gates. The Hall-bar-shaped gate is hidden and is shown by the dashed line in the central part of the sample. (b) View on the central part of the sample on an expanded scale. The Hall-bar width is equal to 50~$\mu$m and the distance between the potential probes is equal to 150~$\mu$m. (c) Schematics of the layer growth sequence and the cross section of the double-gate sample.}
\label{fig1}
\end{figure}

\begin{figure}
\scalebox{0.45}{\includegraphics[angle=0]{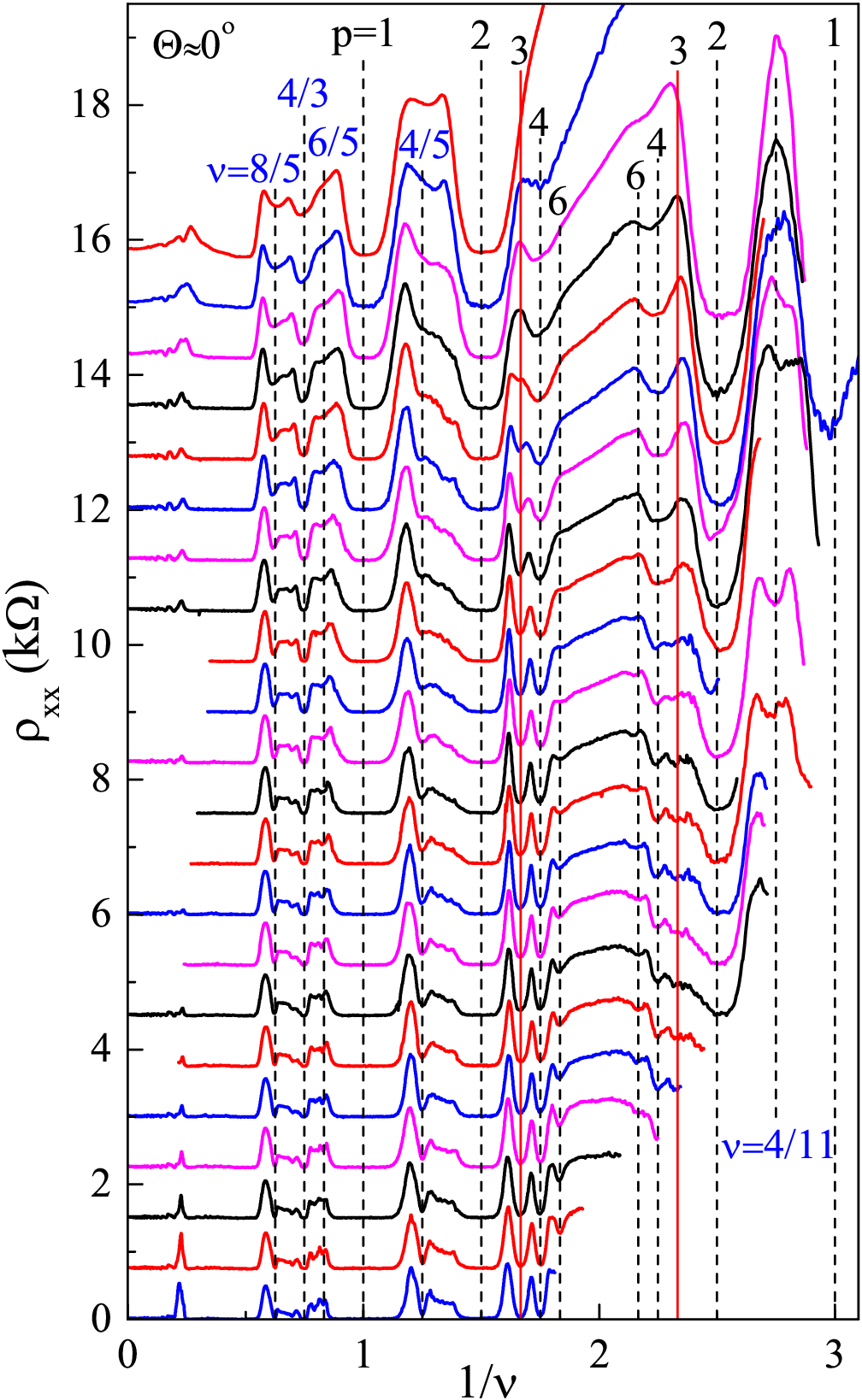}}
\caption{Magnetoresistance of single-gate sample 1 at $T\approx 0.03$~K in perpendicular magnetic fields at electron densities (from top to bottom) 3.19, 3.81, 4.42, 5.03, 5.65, 6.26, 6.88, 7.49, 8.10, 8.72, 9.33, 9.95, 10.6, 11.2, 11.8, 12.4, 13.0, 13.6, 14.9, 16.1, 17.3, and $18.5\times 10^{10}$~cm$^{-2}$. Curves are vertically shifted by 750~$\Omega$ for clarity. Dashed vertical lines mark the expected positions of the observed minima of the resistance, and solid vertical lines correspond to the minima that are expected, but not observed at low densities.}
\label{fig2}
\end{figure}

The samples used were ultrahigh mobility SiGe/Si/SiGe quantum wells similar to those described in Ref.~\cite{melnikov2015ultra}. The peak electron mobility in these samples reaches $\approx 2\times 10^6$~cm$^2$/Vs. The approximately 15~nm wide silicon (001) quantum well is sandwiched between Si$_{0.8}$Ge$_{0.2}$ potential barriers. Contacts to the 2D layer consisted of $\approx 300$~nm Au$_{0.99}$Sb$_{0.01}$ alloy deposited in a thermal evaporator and annealed for 3--5 minutes at 440~$^\circ$C in N$_2$ atmosphere. A 200--300-nm-thick layer of SiO was deposited on the surface of the wafer in a thermal evaporator, and a $>$20-nm-thick aluminum gate was deposited on top of SiO. The samples were patterned in Hall-bar shapes with the distance between the potential probes of 150~$\mu$m and width of 50~$\mu$m using standard photo-lithography. The long side of the Hall bar corresponded to the direction of current parallel to the [110] or [$\bar1$10] crystallographic axis. In tilted magnetic fields, the parallel field component was always perpendicular to the current in order to exclude the influence of ridges on the quantum well surface on the measured resistance \cite{melnikov2017unusual}. In addition to single-gate samples, double-gate samples with a NiCr/Al Hall-bar gate and an Al contact gate were used to minimize the contact resistance at low electron densities in the main part of the sample (Fig.~\ref{fig1}). For making the Hall-bar gate, a 240-nm-thick SiO layer was deposited on the surface of the wafer in a thermal evaporator, and a 60-nm-thick aluminum gate was deposited on top of SiO. In addition, a 20-nm-thick layer of NiCr was deposited on top of Al for better adhesion of subsequent layers. After that, the contact gate was fabricated, for which the whole structure was covered by a 60-nm-thick SiO layer, and a 30-nm-thick aluminum gate was deposited on top of SiO. The contact gate allows for maintaining a high electron density $\approx 2\times10^{11}$~cm$^{-2}$ near the contacts regardless of its value in the main part of the sample. The electron density was controlled by applying a positive dc voltage to the gate relative to the contacts. To improve the quality of contacts and increase the electron mobility, a saturating infrared illumination of the samples was used \cite{melnikov2015ultra,melnikov2017unusual}. The range of accessible electron densities was restricted because of a tunneling of the electrons through the SiGe barrier at high gate voltages \cite{lu2011upper}, whereas the contact resistance increased drastically at very low electron densities/high magnetic fields. Measurements were carried out in an Oxford TLM-400 dilution refrigerator at a temperature $T\approx 0.03$~K. Magnetoresistance was measured with a standard four-terminal lock-in technique in a frequency range 0.2--11~Hz in the linear response regime (the measuring current was in the range between 0.05 and 2~nA). To change the tilt angle of the magnetic field, the sample was warmed up, permanently fixed at a new angle, and cooled down again. The sample state was reproducible in different cool-downs, as inferred from the same dependence, within the experimental uncertainty, of the resistance on electron density in zero magnetic field. Similar results were obtained on three samples of the same wafer.

\begin{figure}
\scalebox{0.435}{\includegraphics[angle=0]{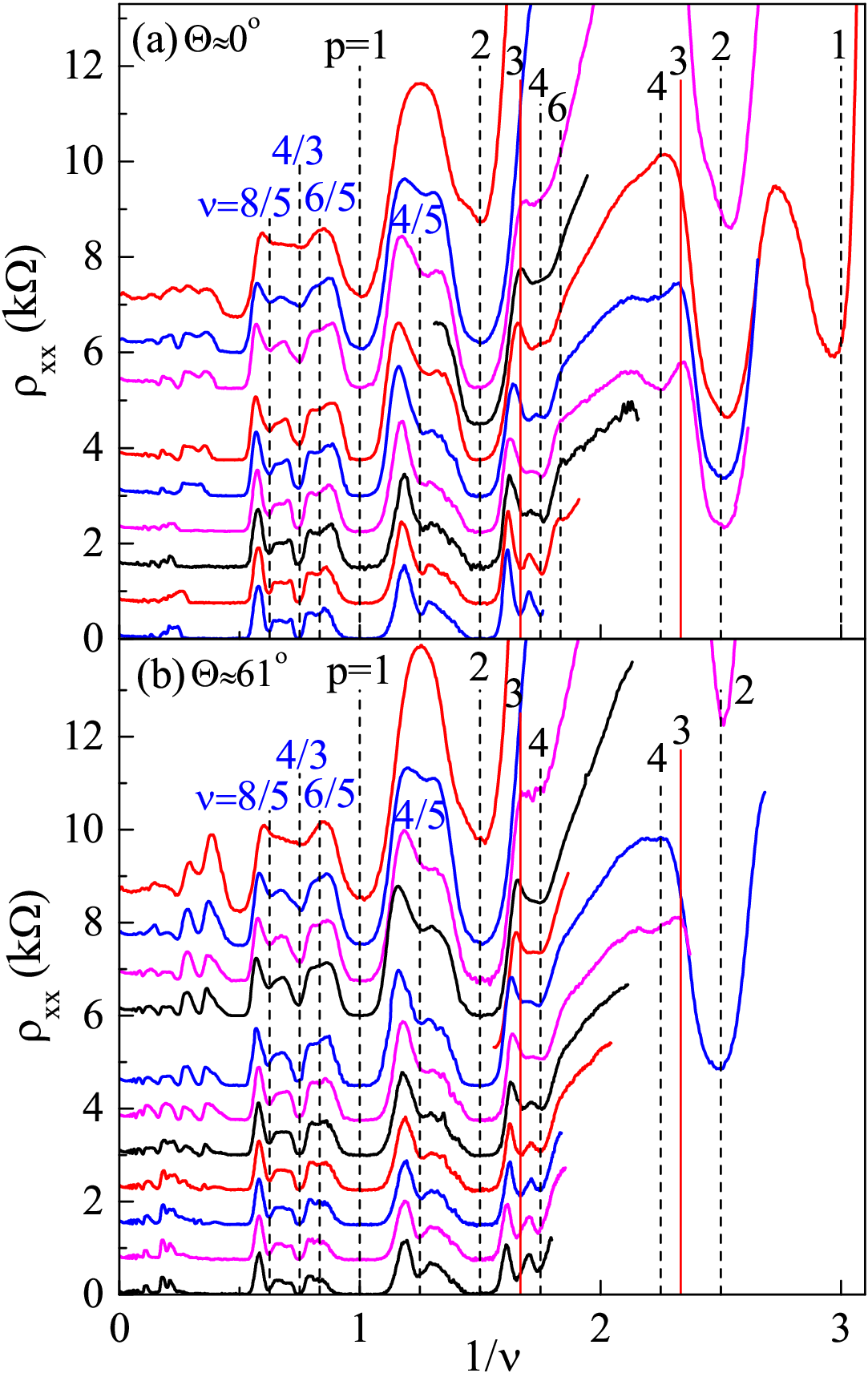}}
\caption{Magnetoresistance of double-gate sample 2 at $T\approx 0.03$~K (a) in perpendicular magnetic fields at electron densities (from top to bottom) 2.14, 2.81, 3.48, 3.81, 4.15, 4.82, 5.49, 6.15, 6.82, and $7.49\times 10^{10}$~cm$^{-2}$, and (b) in tilted magnetic fields at electron densities (from top to bottom) 2.14, 2.81, 3.48, 4.15, 4.48, 4.82, 5.49, 6.15, 6.82, 7.49, 8.16, and $8.83\times 10^{10}$~cm$^{-2}$. Curves are vertically shifted by 750~$\Omega$ for clarity. Dashed vertical lines mark the expected positions of the observed minima of the resistance, and solid vertical lines correspond to the minima that are expected, but not observed at low densities.}
\label{fig3}
\end{figure}

\begin{figure}
\scalebox{0.4}{\includegraphics[angle=0]{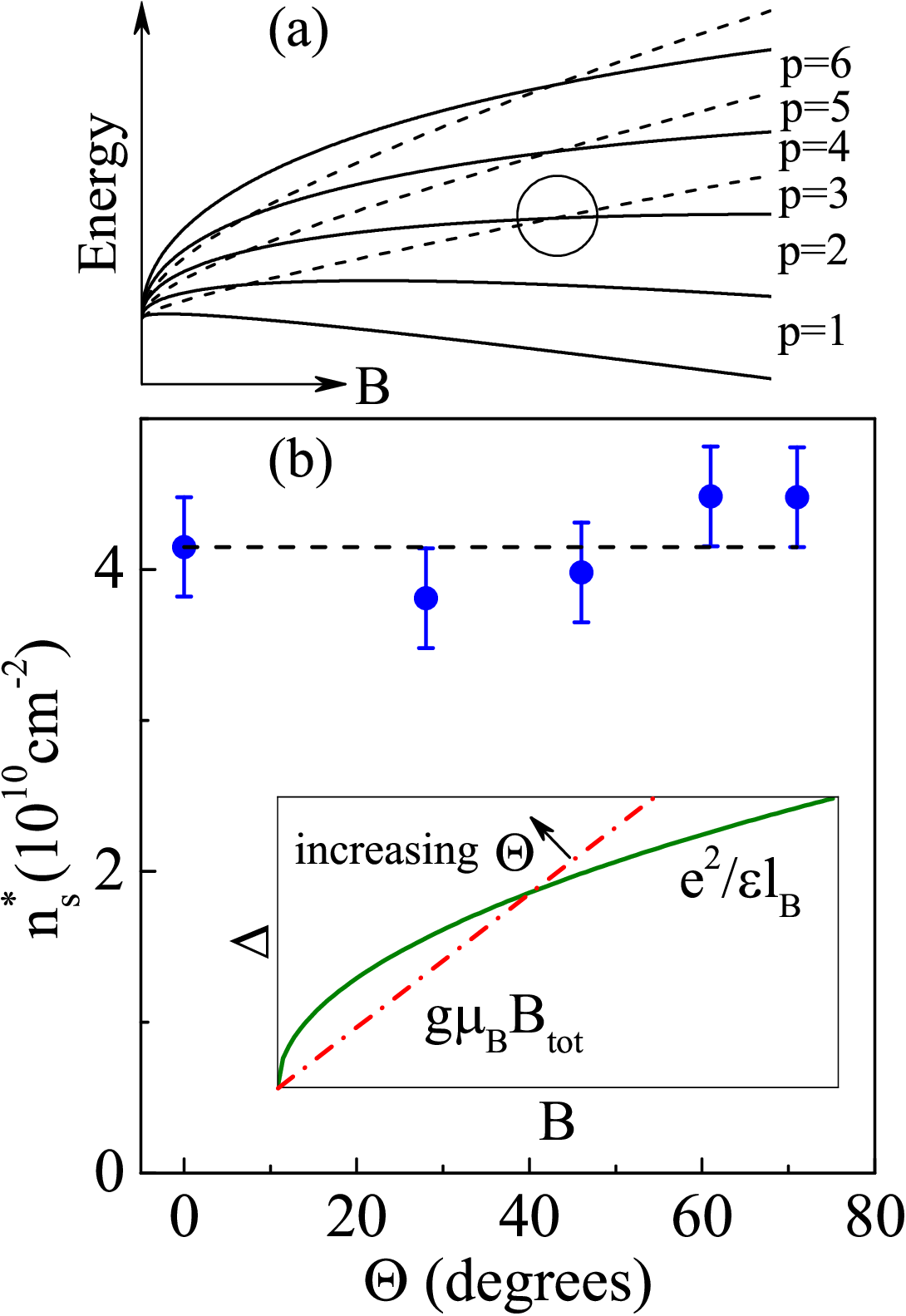}}
\caption{(a) The schematic behavior of the composite fermion levels taking account of the splitting between upper (dashed lines) and lower (solid lines) valleys with changing magnetic field $B$ at fixed $p$. In the region of interest at $p=3$, indicated by the circle, there occurs either simple crossing of the levels or merging/locking of the levels accompanied with a gradual change in the fillings of both levels. (b) The onset density $n_{\text s}^*$ of the resistance minimum at $\nu=3/5$ as a function of the tilt angle in double-gate sample 2. The dashed horizontal line is a fit to the data. The inset schematically (up to a numerical factor) shows the cyclotron energy of composite fermions (the solid line) and the Zeeman energy (the dotted-dashed line) as a function of magnetic field $B$ at a fixed tilt angle. The slope of the straight line increases with increasing $\Theta$.}
\label{fig4}
\end{figure}

The longitudinal resistivity $\rho_{\text {xx}}$ as a function of the inverse filling factor is shown for different electron densities in sample 1 in Fig.~\ref{fig2}. The resistance minima are seen at composite fermion quantum numbers $p=1$, 2, 3, 4, and 6 near $\nu=1/2$ in positive and negative effective fields $B^*$, revealing the high quality of the sample. This is confirmed by the presence of the $\nu=4/5$ and $\nu=4/11$ fractions, corresponding to $p=4/3$, which can be described in terms of the second generation of composite fermions. The behavior of the minima at $p=3$ is unusual in that they disappear below a certain electron density $n_{\text s}^*$, although the surrounding minima at $p=2$ and $p=4$ persist to significantly lower densities. Such a behavior is observed in all samples studied [see, \textit{e.g.}, the data for $\rho_{\text {xx}}$ \textit{vs} $1/\nu$ in sample 2 in Fig.~\ref{fig3}(a)]. Clearly, the prominence of the minima at $p=3$ at low electron densities cannot be explained by level broadening. On the other hand, this finding is strikingly similar to the effect of the disappearance of the cyclotron minima in the magnetoresistance at low electron densities in Si MOSFETs while the spin minima survive down to appreciably lower densities \cite{kravchenko2000shubnikov}, which signifies that the cyclotron splitting becomes equal to the sum of the spin and valley splittings, and the corresponding valley sublevels merge \cite{shashkin2014merging}. Thus, the crossing or merging of both spin and valley sublevels of composite fermions are possible for our case.

We make measurements in tilted magnetic fields in order to distinguish between the spin or valley origin of the effect. The magnetoresistance as a function of the inverse filling factor is shown for the tilt angle $\Theta\approx 61^\circ$ at different electron densities in sample 2 in Fig.~\ref{fig3}(b). Here, we focus on the resistance minimum at $\nu=3/5$. The behavior observed for the $\nu=3/5$ minimum is very similar to that in perpendicular magnetic fields, which holds for all samples and tilt angles. We determine the onset $n_{\text s}^*$ for the $\nu=3/5$ minimum and plot it versus the tilt angle, as shown in Fig.~\ref{fig4}(b). The value $n_{\text s}^*$ turns out to be independent, within the experimental uncertainty, of the tilt angle of the magnetic field. Since the spin splitting is determined by total magnetic field, $\Delta_{\text s}=g\mu_{\text B}B_{\text{tot}}$ (where $g$ is the Land\'e $g$-factor and $\mu_{\text B}$ is the Bohr magneton), one expects that the onset $n_{\text s}^*$ should decrease with the tilt angle [the inset to Fig.~\ref{fig4}(b)], which is in contradiction with the experiment. We arrive at a conclusion that the spin origin of the effect can be excluded, revealing its valley origin. The valley splitting $\Delta_{\text v}$ is expected to be insensitive to the parallel component of the magnetic field \cite{khrapai2003direct} so that the value $n_{\text s}^*$ should be independent of the tilt angle, which is consistent with the experiment. Thus, our results indicate the intersection or merging of the quantum levels of composite fermions with different valley indices. Note that somewhat different values of $n_{\text s}^*\approx 5.3\times 10^{10}$~cm$^{-2}$ and $n_{\text s}^*\approx 4.2\times 10^{10}$~cm$^{-2}$ in samples 1 and 2, correspondingly, can be caused by different values of the valley splitting, as was observed in Ref.~\cite{pudalov2001valley}.

It is clear that for the occurrence of the crossing or merging of the levels of composite fermions with different valley indices, the functional dependences of both splittings on magnetic field $B$ (or electron density) at fixed $p$ should be different. Indeed, the cyclotron energy of composite fermions $\hbar\omega_{\text c}^*$ is determined by the Coulomb interaction energy $e^2/\varepsilon l_{\text B}$, and the valley splitting $\Delta_{\text v}$ in a 2D electron system in Si changes linearly with changing magnetic field $B$ (or electron density) \cite{khrapai2003strong}. In high magnetic fields, the valley splitting exceeds strongly the cyclotron energy of composite fermions so that for the case of $p=3$ all three filled levels of composite fermions belong to the same valley [Fig.~\ref{fig4}(a)]. As the magnetic field is decreased at fixed $p$, the lowest level with the opposite valley index should become coincident with the top filled level, leading to vanishing of the energy gap and the disappearance of the resistance minimum at $p=3$. With a further decrease of the magnetic field there should occur either a simple crossing of the levels and reappearance of the gap or merging/locking of the levels accompanied with a gradual change in the fillings of both levels \cite{shashkin2014merging}. In analogy with the merging of the levels with different valley indices found recently at low electron densities in Si MOSFETs, it is very likely that the merging of the composite fermion levels with different valley indices occurs in our samples. In contrast to the simple crossing of quantum levels, the effect of the interaction-caused level merging is very similar to the band flattening at the Fermi level in strongly correlated Fermi systems, which has been indicated in a SiGe/Si/SiGe 2D electron system by the results of Ref.~\cite{melnikov2017indication}.

It is tempting to attribute the dominance of the $\nu=2/5$ fraction over the $\nu=1/3$ fraction, observed earlier in the magnetoresistance at low densities in this electron system \cite{melnikov2019quantum} [see also Figs.~\ref{fig2} and \ref{fig3}(a)], to the valley effects. In this case, the valley splitting is expected to be smaller than the cyclotron energy of composite fermions. In fact, the problem of the gaps in the spectrum is by far complicated, because on one hand, the value $\Delta_{\text v}$ can be interaction enhanced, depending on the filling factor \cite{khrapai2003direct,khrapai2003strong}, and on the other hand, the bivalley 2D electron system is similar to double-layer electron systems where at the filling factor equal to 1 and 2, a number of novel states can form (see, \textit{e.g}., Refs.~\cite{zheng1997spin,pellegrini1998evidence,khrapai2000canted,stern2000dissipationless,sheng2003phase,kellogg2004vanishing,tutuc2004counterflow}). Certainly, theoretical efforts are needed to look into the gap systematics in the spectrum and possible new states.

Concluding, we report studies of the fractional quantum Hall effect in an ultraclean, strongly interacting bivalley 2D electron system in SiGe/Si/SiGe quantum wells. The resistance minima at the filling factor of composite fermions $p=3$ disappear below a certain electron density, while the surrounding minima at $p=2$ and $p=4$ survive at significantly lower densities, and the onset for the resistance minimum at filling factor $\nu=3/5$ is found to be independent of the tilt angle of the magnetic field. These results point to the intersection or merging of the quantum levels of composite fermions with different valley indices, which reveals the valley effect on the fractions. In analogy with Si MOSFETs, it is very likely that the merging of the composite fermion levels with different valley indices occurs in our samples. The observed experimental effects are surprising and warrant further theoretical consideration.

We are grateful to Don Heiman and Mansour Shayegan for useful discussions. The ISSP group was supported by RFBR Grant No.\ 19-02-00196 and a Russian Government contract. The NTU group acknowledges support by the Ministry of Science and Technology, Taiwan (Projects No.\ 110-2634-F-009-027 and No.\ 109-2622-8-002-003). S.V.K. was supported by NSF Grant No.\ 1904051.


\begin{thebibliography}{31}
\expandafter\ifx\csname natexlab\endcsname\relax\def\natexlab#1{#1}\fi
\expandafter\ifx\csname bibnamefont\endcsname\relax
  \def\bibnamefont#1{#1}\fi
\expandafter\ifx\csname bibfnamefont\endcsname\relax
  \def\bibfnamefont#1{#1}\fi
\expandafter\ifx\csname citenamefont\endcsname\relax
  \def\citenamefont#1{#1}\fi
\expandafter\ifx\csname url\endcsname\relax
  \def\url#1{\texttt{#1}}\fi
\expandafter\ifx\csname urlprefix\endcsname\relax\def\urlprefix{URL }\fi
\providecommand{\bibinfo}[2]{#2}
\providecommand{\eprint}[2][]{\url{#2}}

\bibitem[{\citenamefont{Jain}(1989)}]{jain1989composite}
\bibinfo{author}{\bibfnamefont{J.~K.} \bibnamefont{Jain}},
  \bibinfo{journal}{Phys. Rev. Lett.} \textbf{\bibinfo{volume}{63}},
  \bibinfo{pages}{199} (\bibinfo{year}{1989}).

\bibitem[{\citenamefont{Halperin et~al.}(1993)\citenamefont{Halperin, Lee, and
  Read}}]{halperin1993theory}
\bibinfo{author}{\bibfnamefont{B.~I.} \bibnamefont{Halperin}},
  \bibinfo{author}{\bibfnamefont{P.~A.} \bibnamefont{Lee}}, \bibnamefont{and}
  \bibinfo{author}{\bibfnamefont{N.}~\bibnamefont{Read}},
  \bibinfo{journal}{Phys. Rev. B} \textbf{\bibinfo{volume}{47}},
  \bibinfo{pages}{7312} (\bibinfo{year}{1993}).

\bibitem[{\citenamefont{Jain}(1994)}]{jain1994composite}
\bibinfo{author}{\bibfnamefont{J.~K.} \bibnamefont{Jain}},
  \bibinfo{journal}{Science} \textbf{\bibinfo{volume}{266}},
  \bibinfo{pages}{1199} (\bibinfo{year}{1994}).

\bibitem[{\citenamefont{Chakraborty}(2000)}]{chakraborty2000electron}
\bibinfo{author}{\bibfnamefont{T.}~\bibnamefont{Chakraborty}},
  \bibinfo{journal}{Adv. Phys.} \textbf{\bibinfo{volume}{49}},
  \bibinfo{pages}{959} (\bibinfo{year}{2000}).

\bibitem[{\citenamefont{Jain}(2007)}]{jain2007composite}
\bibinfo{author}{\bibfnamefont{J.~K.} \bibnamefont{Jain}},
  \emph{\bibinfo{title}{Composite Fermions}} (\bibinfo{publisher}{Cambridge
  University Press}, \bibinfo{year}{2007}).

\bibitem[{\citenamefont{Lai et~al.}(2004)\citenamefont{Lai, Pan, Tsui, Lyon,
  M\"uhlberger, and Sch\"affler}}]{lai2004two}
\bibinfo{author}{\bibfnamefont{K.}~\bibnamefont{Lai}},
  \bibinfo{author}{\bibfnamefont{W.}~\bibnamefont{Pan}},
  \bibinfo{author}{\bibfnamefont{D.~C.} \bibnamefont{Tsui}},
  \bibinfo{author}{\bibfnamefont{S.}~\bibnamefont{Lyon}},
  \bibinfo{author}{\bibfnamefont{M.}~\bibnamefont{M\"uhlberger}},
  \bibnamefont{and}
  \bibinfo{author}{\bibfnamefont{F.}~\bibnamefont{Sch\"affler}},
  \bibinfo{journal}{Phys. Rev. Lett.} \textbf{\bibinfo{volume}{93}},
  \bibinfo{pages}{156805} (\bibinfo{year}{2004}).

\bibitem[{\citenamefont{Lu et~al.}(2009)\citenamefont{Lu, Tsui, Lee, and
  Liu}}]{lu2009observation}
\bibinfo{author}{\bibfnamefont{T.~M.} \bibnamefont{Lu}},
  \bibinfo{author}{\bibfnamefont{D.~C.} \bibnamefont{Tsui}},
  \bibinfo{author}{\bibfnamefont{C.-H.} \bibnamefont{Lee}}, \bibnamefont{and}
  \bibinfo{author}{\bibfnamefont{C.~W.} \bibnamefont{Liu}},
  \bibinfo{journal}{Appl. Phys. Lett.} \textbf{\bibinfo{volume}{94}},
  \bibinfo{pages}{182102} (\bibinfo{year}{2009}).

\bibitem[{\citenamefont{Melnikov et~al.}(2019)\citenamefont{Melnikov, Shashkin,
  Dolgopolov, Zhu, Kravchenko, Huang, and Liu}}]{melnikov2019quantum}
\bibinfo{author}{\bibfnamefont{M.~Y.} \bibnamefont{Melnikov}},
  \bibinfo{author}{\bibfnamefont{A.~A.} \bibnamefont{Shashkin}},
  \bibinfo{author}{\bibfnamefont{V.~T.} \bibnamefont{Dolgopolov}},
  \bibinfo{author}{\bibfnamefont{A.~Y.~X.} \bibnamefont{Zhu}},
  \bibinfo{author}{\bibfnamefont{S.~V.} \bibnamefont{Kravchenko}},
  \bibinfo{author}{\bibfnamefont{S.-H.} \bibnamefont{Huang}}, \bibnamefont{and}
  \bibinfo{author}{\bibfnamefont{C.~W.} \bibnamefont{Liu}},
  \bibinfo{journal}{Phys.\ Rev.\ B} \textbf{\bibinfo{volume}{99}},
  \bibinfo{pages}{081106(R)} (\bibinfo{year}{2019}).

\bibitem[{\citenamefont{Bishop et~al.}(2007)\citenamefont{Bishop, Padmanabhan,
  Vakili, Shkolnikov, De~Poortere, and Shayegan}}]{bishop2007valley}
\bibinfo{author}{\bibfnamefont{N.~C.} \bibnamefont{Bishop}},
  \bibinfo{author}{\bibfnamefont{M.}~\bibnamefont{Padmanabhan}},
  \bibinfo{author}{\bibfnamefont{K.}~\bibnamefont{Vakili}},
  \bibinfo{author}{\bibfnamefont{Y.~P.} \bibnamefont{Shkolnikov}},
  \bibinfo{author}{\bibfnamefont{E.~P.} \bibnamefont{De~Poortere}},
  \bibnamefont{and} \bibinfo{author}{\bibfnamefont{M.}~\bibnamefont{Shayegan}},
  \bibinfo{journal}{Phys. Rev. Lett.} \textbf{\bibinfo{volume}{98}},
  \bibinfo{pages}{266404} (\bibinfo{year}{2007}).

\bibitem[{\citenamefont{Padmanabhan et~al.}(2009)\citenamefont{Padmanabhan,
  Gokmen, and Shayegan}}]{padmanabhan2009density}
\bibinfo{author}{\bibfnamefont{M.}~\bibnamefont{Padmanabhan}},
  \bibinfo{author}{\bibfnamefont{T.}~\bibnamefont{Gokmen}}, \bibnamefont{and}
  \bibinfo{author}{\bibfnamefont{M.}~\bibnamefont{Shayegan}},
  \bibinfo{journal}{Phys. Rev. B} \textbf{\bibinfo{volume}{80}},
  \bibinfo{pages}{035423} (\bibinfo{year}{2009}).

\bibitem[{\citenamefont{Padmanabhan
  et~al.}(2010{\natexlab{a}})\citenamefont{Padmanabhan, Gokmen, and
  Shayegan}}]{padmanabhan2010ferromagnetic}
\bibinfo{author}{\bibfnamefont{M.}~\bibnamefont{Padmanabhan}},
  \bibinfo{author}{\bibfnamefont{T.}~\bibnamefont{Gokmen}}, \bibnamefont{and}
  \bibinfo{author}{\bibfnamefont{M.}~\bibnamefont{Shayegan}},
  \bibinfo{journal}{Phys. Rev. Lett.} \textbf{\bibinfo{volume}{104}},
  \bibinfo{pages}{016805} (\bibinfo{year}{2010}{\natexlab{a}}).

\bibitem[{\citenamefont{Padmanabhan
  et~al.}(2010{\natexlab{b}})\citenamefont{Padmanabhan, Gokmen, and
  Shayegan}}]{padmanabhan2010composite}
\bibinfo{author}{\bibfnamefont{M.}~\bibnamefont{Padmanabhan}},
  \bibinfo{author}{\bibfnamefont{T.}~\bibnamefont{Gokmen}}, \bibnamefont{and}
  \bibinfo{author}{\bibfnamefont{M.}~\bibnamefont{Shayegan}},
  \bibinfo{journal}{Phys. Rev. B} \textbf{\bibinfo{volume}{81}},
  \bibinfo{pages}{113301} (\bibinfo{year}{2010}{\natexlab{b}}).

\bibitem[{\citenamefont{Pan et~al.}(2015)\citenamefont{Pan, Baldwin, West,
  Pfeiffer, and Tsui}}]{pan2015fractional}
\bibinfo{author}{\bibfnamefont{W.}~\bibnamefont{Pan}},
  \bibinfo{author}{\bibfnamefont{K.~W.} \bibnamefont{Baldwin}},
  \bibinfo{author}{\bibfnamefont{K.~W.} \bibnamefont{West}},
  \bibinfo{author}{\bibfnamefont{L.~N.} \bibnamefont{Pfeiffer}},
  \bibnamefont{and} \bibinfo{author}{\bibfnamefont{D.~C.} \bibnamefont{Tsui}},
  \bibinfo{journal}{Phys. Rev. B} \textbf{\bibinfo{volume}{91}},
  \bibinfo{pages}{041301(R)} (\bibinfo{year}{2015}).

\bibitem[{\citenamefont{Dolgopolov et~al.}(2018)\citenamefont{Dolgopolov,
  Melnikov, Shashkin, Huang, Liu, and Kravchenko}}]{dolgopolov2018fractional}
\bibinfo{author}{\bibfnamefont{V.~T.} \bibnamefont{Dolgopolov}},
  \bibinfo{author}{\bibfnamefont{M.~Y.} \bibnamefont{Melnikov}},
  \bibinfo{author}{\bibfnamefont{A.~A.} \bibnamefont{Shashkin}},
  \bibinfo{author}{\bibfnamefont{S.-H.} \bibnamefont{Huang}},
  \bibinfo{author}{\bibfnamefont{C.~W.} \bibnamefont{Liu}}, \bibnamefont{and}
  \bibinfo{author}{\bibfnamefont{S.~V.} \bibnamefont{Kravchenko}},
  \bibinfo{journal}{JETP Lett.} \textbf{\bibinfo{volume}{107}},
  \bibinfo{pages}{794} (\bibinfo{year}{2018}).

\bibitem[{\citenamefont{Park and Jain}(2001)}]{park2001the}
\bibinfo{author}{\bibfnamefont{K.}~\bibnamefont{Park}} \bibnamefont{and}
  \bibinfo{author}{\bibfnamefont{J.}~\bibnamefont{Jain}},
  \bibinfo{journal}{Solid State Commun.} \textbf{\bibinfo{volume}{119}},
  \bibinfo{pages}{291} (\bibinfo{year}{2001}).

\bibitem[{\citenamefont{Kravchenko et~al.}(2000)\citenamefont{Kravchenko,
  Shashkin, Bloore, and Klapwijk}}]{kravchenko2000shubnikov}
\bibinfo{author}{\bibfnamefont{S.~V.} \bibnamefont{Kravchenko}},
  \bibinfo{author}{\bibfnamefont{A.~A.} \bibnamefont{Shashkin}},
  \bibinfo{author}{\bibfnamefont{D.~A.} \bibnamefont{Bloore}},
  \bibnamefont{and} \bibinfo{author}{\bibfnamefont{T.~M.}
  \bibnamefont{Klapwijk}}, \bibinfo{journal}{Solid State Commun.}
  \textbf{\bibinfo{volume}{116}}, \bibinfo{pages}{495} (\bibinfo{year}{2000}).

\bibitem[{\citenamefont{Shashkin et~al.}(2014)\citenamefont{Shashkin,
  Dolgopolov, Clark, Shaginyan, Zverev, and Khodel}}]{shashkin2014merging}
\bibinfo{author}{\bibfnamefont{A.~A.} \bibnamefont{Shashkin}},
  \bibinfo{author}{\bibfnamefont{V.~T.} \bibnamefont{Dolgopolov}},
  \bibinfo{author}{\bibfnamefont{J.~W.} \bibnamefont{Clark}},
  \bibinfo{author}{\bibfnamefont{V.~R.} \bibnamefont{Shaginyan}},
  \bibinfo{author}{\bibfnamefont{M.~V.} \bibnamefont{Zverev}},
  \bibnamefont{and} \bibinfo{author}{\bibfnamefont{V.~A.}
  \bibnamefont{Khodel}}, \bibinfo{journal}{Phys. Rev. Lett.}
  \textbf{\bibinfo{volume}{112}}, \bibinfo{pages}{186402}
  (\bibinfo{year}{2014}).

\bibitem[{\citenamefont{Melnikov
  et~al.}(2017{\natexlab{a}})\citenamefont{Melnikov, Shashkin, Dolgopolov,
  Huang, Liu, and Kravchenko}}]{melnikov2017indication}
\bibinfo{author}{\bibfnamefont{M.~Y.} \bibnamefont{Melnikov}},
  \bibinfo{author}{\bibfnamefont{A.~A.} \bibnamefont{Shashkin}},
  \bibinfo{author}{\bibfnamefont{V.~T.} \bibnamefont{Dolgopolov}},
  \bibinfo{author}{\bibfnamefont{S.-H.} \bibnamefont{Huang}},
  \bibinfo{author}{\bibfnamefont{C.~W.} \bibnamefont{Liu}}, \bibnamefont{and}
  \bibinfo{author}{\bibfnamefont{S.~V.} \bibnamefont{Kravchenko}},
  \bibinfo{journal}{Sci.\ Rep.} \textbf{\bibinfo{volume}{7}},
  \bibinfo{pages}{14539} (\bibinfo{year}{2017}{\natexlab{a}}).

\bibitem[{\citenamefont{Melnikov et~al.}(2015)\citenamefont{Melnikov, Shashkin,
  Dolgopolov, Huang, Liu, and Kravchenko}}]{melnikov2015ultra}
\bibinfo{author}{\bibfnamefont{M.~Y.} \bibnamefont{Melnikov}},
  \bibinfo{author}{\bibfnamefont{A.~A.} \bibnamefont{Shashkin}},
  \bibinfo{author}{\bibfnamefont{V.~T.} \bibnamefont{Dolgopolov}},
  \bibinfo{author}{\bibfnamefont{S.-H.} \bibnamefont{Huang}},
  \bibinfo{author}{\bibfnamefont{C.~W.} \bibnamefont{Liu}}, \bibnamefont{and}
  \bibinfo{author}{\bibfnamefont{S.~V.} \bibnamefont{Kravchenko}},
  \bibinfo{journal}{Appl.\ Phys.\ Lett.} \textbf{\bibinfo{volume}{106}},
  \bibinfo{pages}{092102} (\bibinfo{year}{2015}).

\bibitem[{\citenamefont{Melnikov
  et~al.}(2017{\natexlab{b}})\citenamefont{Melnikov, Dolgopolov, Shashkin,
  Huang, Liu, and Kravchenko}}]{melnikov2017unusual}
\bibinfo{author}{\bibfnamefont{M.~Y.} \bibnamefont{Melnikov}},
  \bibinfo{author}{\bibfnamefont{V.~T.} \bibnamefont{Dolgopolov}},
  \bibinfo{author}{\bibfnamefont{A.~A.} \bibnamefont{Shashkin}},
  \bibinfo{author}{\bibfnamefont{S.-H.} \bibnamefont{Huang}},
  \bibinfo{author}{\bibfnamefont{C.~W.} \bibnamefont{Liu}}, \bibnamefont{and}
  \bibinfo{author}{\bibfnamefont{S.~V.} \bibnamefont{Kravchenko}},
  \bibinfo{journal}{J. Appl. Phys.} \textbf{\bibinfo{volume}{122}},
  \bibinfo{pages}{224301} (\bibinfo{year}{2017}{\natexlab{b}}).

\bibitem[{\citenamefont{Lu et~al.}(2011)\citenamefont{Lu, Lee, Huang, Tsui, and
  Liu}}]{lu2011upper}
\bibinfo{author}{\bibfnamefont{T.~M.} \bibnamefont{Lu}},
  \bibinfo{author}{\bibfnamefont{C.-H.} \bibnamefont{Lee}},
  \bibinfo{author}{\bibfnamefont{S.-H.} \bibnamefont{Huang}},
  \bibinfo{author}{\bibfnamefont{D.~C.} \bibnamefont{Tsui}}, \bibnamefont{and}
  \bibinfo{author}{\bibfnamefont{C.~W.} \bibnamefont{Liu}},
  \bibinfo{journal}{Appl. Phys. Lett.} \textbf{\bibinfo{volume}{99}},
  \bibinfo{pages}{153510} (\bibinfo{year}{2011}).

\bibitem[{\citenamefont{Khrapai
  et~al.}(2003{\natexlab{a}})\citenamefont{Khrapai, Shashkin, and
  Dolgopolov}}]{khrapai2003direct}
\bibinfo{author}{\bibfnamefont{V.~S.} \bibnamefont{Khrapai}},
  \bibinfo{author}{\bibfnamefont{A.~A.} \bibnamefont{Shashkin}},
  \bibnamefont{and} \bibinfo{author}{\bibfnamefont{V.~T.}
  \bibnamefont{Dolgopolov}}, \bibinfo{journal}{Phys. Rev. Lett.}
  \textbf{\bibinfo{volume}{91}}, \bibinfo{pages}{126404}
  (\bibinfo{year}{2003}{\natexlab{a}}).

\bibitem[{\citenamefont{{Pudalov} et~al.}(2001)\citenamefont{{Pudalov},
  {Punnoose}, {Brunthaler}, {Prinz}, and {Bauer}}}]{pudalov2001valley}
\bibinfo{author}{\bibfnamefont{V.~M.} \bibnamefont{{Pudalov}}},
  \bibinfo{author}{\bibfnamefont{A.}~\bibnamefont{{Punnoose}}},
  \bibinfo{author}{\bibfnamefont{G.}~\bibnamefont{{Brunthaler}}},
  \bibinfo{author}{\bibfnamefont{A.}~\bibnamefont{{Prinz}}}, \bibnamefont{and}
  \bibinfo{author}{\bibfnamefont{G.}~\bibnamefont{{Bauer}}},
  \bibinfo{journal}{arXiv e-prints} \bibinfo{eid}{cond-mat/0104347}
  (\bibinfo{year}{2001}).

\bibitem[{\citenamefont{Khrapai
  et~al.}(2003{\natexlab{b}})\citenamefont{Khrapai, Shashkin, and
  Dolgopolov}}]{khrapai2003strong}
\bibinfo{author}{\bibfnamefont{V.~S.} \bibnamefont{Khrapai}},
  \bibinfo{author}{\bibfnamefont{A.~A.} \bibnamefont{Shashkin}},
  \bibnamefont{and} \bibinfo{author}{\bibfnamefont{V.~T.}
  \bibnamefont{Dolgopolov}}, \bibinfo{journal}{Phys. Rev. B}
  \textbf{\bibinfo{volume}{67}}, \bibinfo{pages}{113305}
  (\bibinfo{year}{2003}{\natexlab{b}}).

\bibitem[{\citenamefont{Zheng et~al.}(1997)\citenamefont{Zheng, Radtke, and
  Das~Sarma}}]{zheng1997spin}
\bibinfo{author}{\bibfnamefont{L.}~\bibnamefont{Zheng}},
  \bibinfo{author}{\bibfnamefont{R.~J.} \bibnamefont{Radtke}},
  \bibnamefont{and}
  \bibinfo{author}{\bibfnamefont{S.}~\bibnamefont{Das~Sarma}},
  \bibinfo{journal}{Phys. Rev. Lett.} \textbf{\bibinfo{volume}{78}},
  \bibinfo{pages}{2453} (\bibinfo{year}{1997}).

\bibitem[{\citenamefont{Pellegrini et~al.}(1998)\citenamefont{Pellegrini,
  Pinczuk, Dennis, Plaut, Pfeiffer, and West}}]{pellegrini1998evidence}
\bibinfo{author}{\bibfnamefont{V.}~\bibnamefont{Pellegrini}},
  \bibinfo{author}{\bibfnamefont{A.}~\bibnamefont{Pinczuk}},
  \bibinfo{author}{\bibfnamefont{B.~S.} \bibnamefont{Dennis}},
  \bibinfo{author}{\bibfnamefont{A.~S.} \bibnamefont{Plaut}},
  \bibinfo{author}{\bibfnamefont{L.~N.} \bibnamefont{Pfeiffer}},
  \bibnamefont{and} \bibinfo{author}{\bibfnamefont{K.~W.} \bibnamefont{West}},
  \bibinfo{journal}{Science} \textbf{\bibinfo{volume}{281}},
  \bibinfo{pages}{799} (\bibinfo{year}{1998}).

\bibitem[{\citenamefont{Khrapai et~al.}(2000)\citenamefont{Khrapai, Deviatov,
  Shashkin, Dolgopolov, Hastreiter, Wixforth, Campman, and
  Gossard}}]{khrapai2000canted}
\bibinfo{author}{\bibfnamefont{V.~S.} \bibnamefont{Khrapai}},
  \bibinfo{author}{\bibfnamefont{E.~V.} \bibnamefont{Deviatov}},
  \bibinfo{author}{\bibfnamefont{A.~A.} \bibnamefont{Shashkin}},
  \bibinfo{author}{\bibfnamefont{V.~T.} \bibnamefont{Dolgopolov}},
  \bibinfo{author}{\bibfnamefont{F.}~\bibnamefont{Hastreiter}},
  \bibinfo{author}{\bibfnamefont{A.}~\bibnamefont{Wixforth}},
  \bibinfo{author}{\bibfnamefont{K.~L.} \bibnamefont{Campman}},
  \bibnamefont{and} \bibinfo{author}{\bibfnamefont{A.~C.}
  \bibnamefont{Gossard}}, \bibinfo{journal}{Phys. Rev. Lett.}
  \textbf{\bibinfo{volume}{84}}, \bibinfo{pages}{725} (\bibinfo{year}{2000}).

\bibitem[{\citenamefont{Stern et~al.}(2000)\citenamefont{Stern, Das~Sarma,
  Fisher, and Girvin}}]{stern2000dissipationless}
\bibinfo{author}{\bibfnamefont{A.}~\bibnamefont{Stern}},
  \bibinfo{author}{\bibfnamefont{S.}~\bibnamefont{Das~Sarma}},
  \bibinfo{author}{\bibfnamefont{M.~P.~A.} \bibnamefont{Fisher}},
  \bibnamefont{and} \bibinfo{author}{\bibfnamefont{S.~M.}
  \bibnamefont{Girvin}}, \bibinfo{journal}{Phys. Rev. Lett.}
  \textbf{\bibinfo{volume}{84}}, \bibinfo{pages}{139} (\bibinfo{year}{2000}).

\bibitem[{\citenamefont{Sheng et~al.}(2003)\citenamefont{Sheng, Balents, and
  Wang}}]{sheng2003phase}
\bibinfo{author}{\bibfnamefont{D.~N.} \bibnamefont{Sheng}},
  \bibinfo{author}{\bibfnamefont{L.}~\bibnamefont{Balents}}, \bibnamefont{and}
  \bibinfo{author}{\bibfnamefont{Z.}~\bibnamefont{Wang}},
  \bibinfo{journal}{Phys. Rev. Lett.} \textbf{\bibinfo{volume}{91}},
  \bibinfo{pages}{116802} (\bibinfo{year}{2003}).

\bibitem[{\citenamefont{Kellogg et~al.}(2004)\citenamefont{Kellogg, Eisenstein,
  Pfeiffer, and West}}]{kellogg2004vanishing}
\bibinfo{author}{\bibfnamefont{M.}~\bibnamefont{Kellogg}},
  \bibinfo{author}{\bibfnamefont{J.~P.} \bibnamefont{Eisenstein}},
  \bibinfo{author}{\bibfnamefont{L.~N.} \bibnamefont{Pfeiffer}},
  \bibnamefont{and} \bibinfo{author}{\bibfnamefont{K.~W.} \bibnamefont{West}},
  \bibinfo{journal}{Phys. Rev. Lett.} \textbf{\bibinfo{volume}{93}},
  \bibinfo{pages}{036801} (\bibinfo{year}{2004}).

\bibitem[{\citenamefont{Tutuc et~al.}(2004)\citenamefont{Tutuc, Shayegan, and
  Huse}}]{tutuc2004counterflow}
\bibinfo{author}{\bibfnamefont{E.}~\bibnamefont{Tutuc}},
  \bibinfo{author}{\bibfnamefont{M.}~\bibnamefont{Shayegan}}, \bibnamefont{and}
  \bibinfo{author}{\bibfnamefont{D.~A.} \bibnamefont{Huse}},
  \bibinfo{journal}{Phys. Rev. Lett.} \textbf{\bibinfo{volume}{93}},
  \bibinfo{pages}{036802} (\bibinfo{year}{2004}).

\end{thebibliography}


\end{document}